\documentclass{article}
\usepackage{spconf,amsmath,graphicx}
\usepackage{amsmath,amssymb,amsfonts}
\usepackage{algorithmic}
\usepackage{algorithm}
\usepackage{array}
\usepackage[caption=false,font=normalsize,labelfont=sf,textfont=sf]{subfig}
\usepackage{textcomp}
\usepackage{stfloats}
\usepackage{url}
\usepackage{bm}
\usepackage{verbatim}
\usepackage{graphicx}
\usepackage{cite}
\usepackage{amsthm}
\usepackage{caption}
\usepackage{graphicx}
\usepackage{float} 
\usepackage{subcaption}
\usepackage{hyperref}
\usepackage{gensymb}

\title{Secure Analog Beamforming Design for Wireless Communication Systems With Movable Antennas}
\name{Weijie Xiong$^{\star}$\qquad Kai Zhong$^{\star}$\qquad Zhiling Xiao$^{\star}$\qquad Jingran Lin$^{\star \dagger}$ \qquad Qiang Li$^{\star \dagger}$ \thanks{This work was supported in part by the Natural Science Foundation of China (NSFC) under Grant 62171110. \textit{(Corresponding author: Jingran Lin.)}.}}
\address{$^{\star}$ University of Electronic Science and Technology of China, Chengdu 611731, China\\$^{\dagger}$ Laboratory of Electromagnetic Space Cognition and Intelligent Control, Beijing 100083, China}

\begin{document}
\ninept
\maketitle

\begin{abstract}
Movable antennas (MA) are a novel technology that allows for the flexible adjustment of antenna positions within a specified region, thereby enhancing the performance of wireless communication systems. In this paper, we explore the use of MA to improve physical layer security in an analog beamforming (AB) communication system. Our goal is to maximize the secrecy rate by jointly optimizing the transmit AB and MA position, subject to constant modulus (CM) constraints on the AB and position constraints for the MA. The resulting problem is non-convex, and we propose a penalty product manifold (PPM) method to solve it efficiently. Specifically, we convert the inequality constraints related to MA position into a penalty function using smoothing techniques, thereby reformulating the problem as an unconstrained optimization on the product manifold space (PMS). We then derive a parallel conjugate gradient descent (PCGD) algorithm to update both the AB and MA position on the PMS. This method is efficient, providing an analytical solution at each step and ensuring convergence to a KKT point. Simulation results show that the MA system achieves a higher secrecy rate than systems with fixed-position antennas.
\end{abstract}
\section{Introduction}
In conventional physical-layer security (PLS) systems, analog beamforming (AB) enhances communication security by adjusting the phase shifts of hardware-efficient analog phase shifters (PS) to selectively strengthen or weaken signal reception \cite{cheng2024massive, li2019constant, li2017constant}. However, existing AB techniques rely on fixed-position antennas (FPAs), where all antennas are deployed at static locations. This fixed configuration poses challenges in maintaining satisfactory PLS performance, particularly in more complex communication environments where the channels of eavesdroppers and legitimate receivers exhibit strong correlation \cite{xiao2023array, hu2024secure, cheng2024enabling}.

To address this issue, movable antennas (MA) have been proposed, introducing an additional spatial degree of freedom (DoF) that allows flexible adjustment of antenna positions within a specified region. This flexibility provides an effective solution for improving system performance in complex environments \cite{zhu2023movable, zheng2024flexible, chen2023joint}. By enabling the modification of steering vectors at different angles, MA allows reconfiguration of wireless channels, thereby enhancing communication capacity. Previous studies have demonstrated the significant potential of MA-enabled communications. For example, applications, hardware architectures, and performance benefits of MA were discussed in \cite{ma2023compressed, xiao2024channel}. Compressed sensing-based channel estimation for MA is presented in \cite{zhu2023modeling, zhu2023movable2}, while joint MIMO designs were in \cite{ma2024multi}. Channel modeling and outage analysis for MA was explored in \cite{pi2023multiuser}, and MA’s impact on signal power enhancement/nulling was investigated in \cite{yang2024flexible}.

More specifically, MA technolog has also been explored to enhance PLS \cite{hu2024secure,cheng2024enabling}. However, these PLS designs focus on fully digital beamforming (FDB) at the transmitter, incurring high hardware costs due to the need for a dedicated RF chain for each antenna. In contrast, AB systems require only PS at the transmitter, and thus offer a more hardware-efficient solution. Nevertheless, AB systems face challenges due to constant modulus (CM) signaling. While FDB systems can achieve optimal secure beamforming under the total transmit power constraint with a closed-form solution via generalized eigendecomposition \cite{5485016}, this approach is not feasible for AB systems. The generalized eigenvector is no longer optimal, and a simple projection onto the CM constraints leads to significant secrecy rate loss. Therefore, the methods in \cite{hu2024secure,cheng2024enabling} are not applicable to AB designs.

To address these challenges, we note that the complex circle manifold (CCM) naturally satisfies the CM constraints in AB systems. Additionally, since the inequality MA position constraints must be strictly enforced, we use the penalty function method to penalize any violations. Building on these insights, we propose the penalty product manifold (PPM) method. Specifically, we convert the inequality constraints into a penalty function using smoothing techniques, thereby reformulating the problem as an unconstrained optimization on the product manifold space (PMS). We then derive a parallel conjugate gradient descent (PCGD) algorithm to update both the AB and MA position on the PMS. This method is highly efficient, providing a simple analytical solution at each step and ensuring convergence to a KKT point. Simulation results show that the MA system achieves a higher secrecy rate than systems with fixed-position, sparse-selection, and random-selection antennas.

\section{System Model and Problem Formulation}
\label{sec:format}
Consider a network where the transmitter (Alice) uses an analog beamformer to broadcast confidential information to a legitimate receiver (Bob), in the presence of an eavesdropper (Eve) who attempts to intercept the data sent from Alice to Bob. Bob and Eve are equipped with a single, fixed-position antenna, while Alice is equipped with a linear MA array of size $M$. The position of the $m$-th antenna at Alice is denoted by ${p_m}, 1 \le m \le M$. The positions of the $M$ antennas can then be compactly expressed as ${\bf p}=[p_1,p_2,...,p_M]^T \in \mathbb{R}^{M}$, where $(\cdot)^T$ is the transpose operation. Let $\mathbf{w} \in \mathbb{C}^{M}$ be the transmit analog beamforming (AB) controlled by the PS with constant modulus (CM), i.e.,
\begin{equation}
|w_i| = \sqrt {{P}},\quad \forall i = 1,...,M, \label{beamformingw}
\end{equation}
where $|\cdot|$ is the absolute value operator; $ \sqrt {{P}} > 0$ represents the per-antenna transmit power. Then, the received signal at Bob and Eve are respectively given by,
\begin{subequations}
\begin{align}
& {y}_b(t)=\mathbf{h}_{b}^H \mathbf{w}s(t) +{{n}}_b(t) \in \mathbb{C}, \\
& {y}_e(t)= {\bf{h}}_{e}^H {\bf{w}}s(t) +{{n}}_e(t) \in \mathbb{C},
\end{align}
\end{subequations}
where $(\cdot)^H$ denotes Hermitian transpose operation; $s(t) \in \mathbb{C}$ is the coded confidential information for Bob with unit power; ${{n}}_b(t) \sim \mathcal{C N}({0}, \sigma_b^2)$ and ${{n}}_e(t) \sim \mathcal{C N}({0}, \sigma_e^2)$ are Gaussian noises; $\mathbf{h}_{b}$ and ${\bf{h}}_{e}$ denote the channels from the transmitter to Bob and to Eve respectively. It is worth noting that the MA-based channel vectors are
determined by the signal propagation environment and the positions of MAs. We consider the field-response based channel model given by \cite{ma2023mimo},
\begin{equation}
{\bf h}_i = \sqrt{\frac{1}{L_i}}\sum_{l=1}^{L_i}\beta_{i,l}{\bf a}(\theta_{i,l},{\bf p}), \quad i \in \{b,e \},
\label{channel}
\end{equation}
where $\beta_{i,l}$ is the complex path gain for the $l$-th path; $L_i$ is number of spatial channel paths; ${\bf a}(\theta_{i,l},{\bf p})$ is the far-field steering vector given as,
\begin{equation}
    {\bf a}(\theta_{i,l},{\bf p})=[e^{j\tfrac{2\pi }{\lambda }\text{cos}\theta_{i,l} p_1 },e^{j\tfrac{2\pi}{\lambda }\text{cos}\theta_{i,l} p_2 },...,e^{j\tfrac{2\pi}{\lambda }\text{cos}\theta_{i,l} p_{L} }]^T
\end{equation}
where $\lambda$ is the wavelength; $\theta_{i,l}$ is the azimuth angles of the $l$-th path. From (\ref{beamformingw})-(\ref{channel}), the secrecy rate is expressed as \cite{Khisti2007SecureTW},
\begin{equation}
{R_s({\bf w},{\bf p})}=\left[\begin{array}{l} \log(1+\frac{\sum_{l=1}^{L_b}|\beta_{b,l}|^2| {\bf a}(\theta_{b,l},{\bf p})^H \mathbf{w}|^2}{L_b \sigma_b^2}) \\  - \log(1+\frac{\sum_{l=1}^{L_e}|\beta_{e,l}|^2| {\bf a}(\theta_{e,l},{\bf p})^H \mathbf{w}|^2}{L_e \sigma_e^2}) \end{array}\right]^{+},
\end{equation}
where $[\cdot]^+=\text{max}\{0,\cdot\}$. Our goal is to maximize ${R_s({\bf w},{\bf p})}$, by jointly optimizing the positions of MAs $\bf p$ and the analog transmit beamformer $\bf w$ at Alice. Hence, the optimization problem is formulated as,
\begin{subequations}
\begin{align}
& \max _{{\bf w}, {\bf p}} \quad \left[\begin{array}{l} \log(1+\frac{P\sum_{l=1}^{L_b}|\beta_{b,l}|^2| {\bf a}(\theta_{b,l},{\bf p})^H \mathbf{w}|^2}{L_b \sigma_b^2}) \\  - \log(1+\frac{P\sum_{l=1}^{L_e}|\beta_{e,l}|^2| {\bf a}(\theta_{e,l},{\bf p})^H \mathbf{w}|^2}{L_e \sigma_e^2}) \end{array}\right]^{+}, \label{Rssecret} \\
& \text { s.t. } \quad\left|w_i\right|=1, \quad i=1, \ldots, M, \label{CMC} \\
&  \quad\quad\quad {p_{m+1}} - {p_{m}} \ge \frac{\lambda}{2}, \quad 1 \le m \le M-1, \label{minidistance}\\
&  \quad\quad\quad {p_{1}} \ge 0, \quad {p_{M}} \le L, \label{antennaselect}
\end{align}
\label{objectivefunction}
\end{subequations}
We have normalized the magnitude of the beamformer element-wise to one by multiplying the channels by $\sqrt{P}$; constraints (\ref{minidistance}) ensure that the distance between any two MAs is no smaller than $\tfrac{\lambda}{2}$ tto avoid the coupling effect; constraints (\ref{antennaselect}) guarantees that the position of any MA is no greater than $L$ and no smaller than zero. Additionally, to ensure that (\ref{minidistance}) always holds, it is evident that $L$ in (\ref{antennaselect}) should be no smaller than $\frac{\lambda}{2}(M - 1)$. 

To avoid directly optimizing the maximum objective function (\ref{objectivefunction}), we drop the logarithmic operation and exchange the numerator and denominator. Thus, problem (\ref{objectivefunction}) can be equivalently rewritten as,
\begin{equation}
\begin{aligned}
& \min _{{\bf w}, {\bf p}} \quad {\hat{R}({\bf w},{\bf p})}=\frac{1+t_e\sum_{l=1}^{L_e}|\beta_{e,l}|^2| {\bf a}(\theta_{e,l},{\bf p})^H \mathbf{w}|^2}{1+t_b\sum_{l=1}^{L_b}|\beta_{b,l}|^2| {\bf a}(\theta_{b,l},{\bf p})^H \mathbf{w}|^2},  \\
& \text { s.t. \quad(\ref{CMC}), (\ref{minidistance}), (\ref{antennaselect}) are satisfied },
\end{aligned}
\label{refornewobjuse}
\end{equation}
where $t_e = \frac{P}{L_e\sigma_e^2}$ and $t_b = \frac{P}{L_b\sigma_b^2}$ are constants for symbol simplification.

Generally, problem (\ref{refornewobjuse}) is non-convex due to the following facts: the concave objective function, the CM constraints (\ref{CMC}), and the coupled-variable constraints (\ref{minidistance}). To address these challenges, we note that the CCM naturally satisfies the CM constraints (\ref{CMC}). Additionally, since the coupled-variable constraints (\ref{minidistance}) must be strictly enforced, we use the penalty function method to penalize violations, ensuring compliance with these constraints. Building on these insights, we propose the penalty product manifold (PPM) method.

\section{The Proposed PPM Method}
In the following, we provide detailed expressions for solving problem (\ref{refornewobjuse}) using the PPM method. Specifically, we first convert the inequality position constraints related to MA position into a penalty function using smoothing techniques, reformulating the problem as an unconstrained optimization on the PMS. Then, we derive a PCGD algorithm to update the AB and MA position over the PMS.

\subsection{Problem Reformulation}
To address the challenge of the coupled-variable inequality constraints (\ref{minidistance}), we incorporate them into the objective function by penalizing the violations. This transforms the problem into an exterior penalty function optimization with CM constraints, given as,
 \begin{subequations}
\begin{align}
& \min _{{\bf w}, {\bf p}} \quad {\hat{R}({\bf w},{\bf p})}  + \rho \left[\begin{array}{l} \textstyle\sum_{m=1}^{M-1}   
 \max\{0,g_m(p_{m+1},p_{m})\} \\\quad\quad+ \max\{0,f(p_{1})\}\\\quad\quad+ \max\{0,f(p_{M})\} \end{array}\right], \label{reforobj2} \\
& \text { s.t. } \quad\left|w_i\right|=1, \quad i=1, \ldots, M,
\end{align}
\label{refor2}
\end{subequations}
where $\rho>0$ is the penalty factor and,
 \begin{subequations}
\begin{align}
& g_m(p_{m+1},p_{m}) = {p_m} - {p_{m+1}} + \frac{\lambda}{2},\\
& f(p_{1}) = -{p_1}, \quad f(p_{M}) = {p_M} - L.
\end{align}
\end{subequations}

Note that (\ref{refor2}) is non-smooth and challenging to solve directly. However, the two-term maximums in the cost function (\ref{refor2}) can be smoothed using the LogSumExp function \cite{blanchard2021accurately}, denoted as $\max\{a,b\} \approx \log(e^{a}+e^{b})$. Thus, (\ref{refor2}) can be transformed into a smooth penalty function optimization problem, given as,
\begin{subequations}
\begin{align}
& \min _{{\bf w}, {\bf p}}   \phi({\bf w}, {\bf p})= \left[\begin{array}{l}{\quad\quad\quad\quad\hat{R}({\bf w},{\bf p})} \\+ \rho  \sum_{m=1}^{M-1}   
 \log(1+e^{g_m(p_{m+1},p_{m})}) \\ \quad\quad\quad +  \rho\log(1+e^{f(p_{1})}) \\\quad\quad\quad+ \rho\log(1+e^{f(p_{M})}) \end{array}\right], \label{reforobj2} \\
& \text { s.t. }\left|w_i\right|=1, \quad i=1, \ldots, M.\label{constantmodulrefor2}
\end{align}
\label{refor3}
\end{subequations}

Note that the CM constraints in (\ref{refor3}) are difficult to express and handle in a linear space. Manifolds, being flexible and capable of capturing non-linear relationships, provide a natural framework for expressing and incorporating these constraints. The local Euclidean nature of the manifold allows complex relationships to be expressed in a way that aligns with the underlying geometry.

\subsection{Manifold Space Construction}
In this subsection, we construct a PMS designed to satisfy the constraints in (\ref{refor3}). It is essential to ensure that the PMS adheres to the specified constraints of CM beamforming $\bf w$ and MA position $\bf p$. Consistency with these constraints is crucial for the effectiveness and reliability of the PMS.

Constraints (\ref{constantmodulrefor2}) satisﬁes the complex circle manifold ${\cal M}_{\bf w}$ \cite{an2024robust},
\begin{equation}
\mathcal{M}_{\bf w}=\left\{{\bf w} \in \mathbb{C}^{M} | \,|w_i| =1, \quad i=1, \ldots, M \right\}. \label{manifoldw}
\end{equation}
Variable $\bf p$ satisﬁes the Euclidean space manifold ${\cal M}_{\bf p}$,
\begin{equation}
\mathcal{M}_{\bf p}=\left\{{\bf p} \in \mathbb{R}^{M} \right\}. \label{manifoldf}
\end{equation}

To represent the feasible set of solutions, we construct the PMS as the Cartesian product of those individual manifolds in (\ref{manifoldw}) and (\ref{manifoldf}) \cite{li2020riemannian}. The PMS is given by,
\begin{equation}
\mathcal{M}=\mathcal{M}_{\bf w} \times \mathcal{M}_{\bf p}=\left\{({\bf w}, {\bf p}): |w_i| =1, \forall i, {\bf p} \in \mathbb{R}^{M} \right\}.   \label{productmanifold} 
\end{equation}

Based on (\ref{productmanifold}), problem (\ref{refor3}) can be reformulated as an unconstrained problem on the PMS, given as,
\begin{equation}
\min _{({\bf w}, {\bf p})\in\mathcal{M}} \phi({\bf w}, {\bf p}).
\label{reforoverpro}    
\end{equation}

\subsection{The PCGD Algorithm for Solving (\ref{reforoverpro})}
To solve (\ref{reforoverpro}), we derive the PCGD algorithm on the PMS. The process of this algorithm can be divided into three steps: 1) Calculate the descent direction; 2) Calculate the descending step size; 3) Update iteration point.

\subsubsection{Calculation of the descent direction}
At $l$-th iteration of the algorithm, the descent direction is given as,
\begin{equation}
\begin{aligned}
{\bf d}_l & = [{\bf d}_{{\bf w}_l};{\bf d}_{{\bf p}_l}] \\& = -\text{grad}(\phi({\bf w}_l, {\bf p}_l)) + \sigma_{l} \text{Trans}_{({\bf w}_{l},{\bf p}_{l}) \leftarrow ({\bf w}_{l-1},{\bf p}_{l-1}) }({\bf d}_{l-1}),
\end{aligned} 
\label{Calculationdescent}
\end{equation}
where,
\begin{itemize}
\item $\text{grad}(\phi({\bf w}, {\bf p}))$ is the Riemannian gradient of the objective function $\phi({\bf w}, {\bf p})$ on the manifold tangent space given as  \cite{boumal2023intromanifolds},
\begin{equation}
\text{grad}(\phi({\bf w}, {\bf p})) = [\text{grad}(\phi_{\bf w}({\bf w}, {\bf p}));\text{grad}(\phi_{\bf p}({\bf w}, {\bf p}))],
\end{equation}
where,
\begin{equation}
\begin{aligned}
&\text{grad}(\phi_{\bf w}({\bf w}, {\bf p})) = \nabla_{\bf w} \phi({\bf w}, {\bf p})  \\ &\quad\quad\quad\quad\quad\quad\quad\quad - \Re(\nabla_{\bf w} \phi({\bf w}, {\bf p})  \odot {\bf w}^*) \odot {\bf w}, \\ 
&\text{grad}(\phi_{\bf p}({\bf w}, {\bf p})) = \nabla_{\bf p} \phi({\bf w}, {\bf p}),
\end{aligned}
\label{updategradient}
\end{equation}
where $\nabla \phi({\bf w}, {\bf p}) = [\nabla_{\bf w} \phi({\bf w}, {\bf p}); \nabla_{\bf p} \phi({\bf w}, {\bf p}) ]$  is the Euclidean gradient of the objective function; $\odot$ denotes elementwise product. Detailed derivation of $\nabla \phi({\bf w}, {\bf p})$ are omitted due to space; readers can refer to \cite{hu2024secure} for more information.

\item $\text{Trans}_{({\bf w}_{l},{\bf p}_{l}) \leftarrow ({\bf w}_{l-1},{\bf p}_{l-1}) }({\bf d}_{l-1})$ is the transportation operation to relocate the vector ${\bf d}_{l-1}$ from the point $({\bf w}_{l-1},{\bf p}_{l-1})\in\mathcal{M}$ to an alternate point $({\bf w}_{l},{\bf p}_{l})\in\mathcal{M}$, given as \cite{boumal2023intromanifolds},
\begin{equation}
\begin{aligned}
   & \text{Trans}_{({\bf w}_{l},{\bf p}_{l}) \leftarrow ({\bf w}_{l-1},{\bf p}_{l-1}) }({\bf d}_{l-1}) \\ &\quad\quad\quad\quad\quad\quad\quad= \left[\begin{array}{l} \text{Trans}_{{\bf w}_{l} \leftarrow {\bf w}_{l-1} }({\bf d}_{{\bf w}_{l-1}});\\\text{Trans}_{{\bf p}_{l} \leftarrow {\bf p}_{l-1} }({\bf d}_{{\bf p}_{l-1}})\end{array}\right],
\end{aligned}
\end{equation}
where,
\begin{equation}
\begin{aligned}
&  \text{Trans}_{{\bf w}_{l} \leftarrow {\bf w}_{l-1} }({\bf d}_{{\bf w}_{l-1}}) = {\bf d}_{{\bf w}_{l-1}} - \Re({\bf d}_{{\bf w}_{l-1}}  \odot {\bf w}^*) \odot {\bf w}  ,\\
& \text{Trans}_{{\bf p}_{l} \leftarrow {\bf p}_{l-1} }({\bf d}_{{\bf p}_{l-1}}) = {\bf d}_{{\bf p}_{l-1}}.
\end{aligned}
\end{equation}

\item $\sigma_{l}$ is the Polak–Ribiere parameter determined by \cite{babaie2015hybridization}.

\end{itemize}

\subsubsection{Calculation of the descending step size}
We use the Armijo method \cite{li2019convergence} to update the step size. This method adaptively determines the step size using information from the previous step, thereby accelerating convergence, as given below,
\begin{equation}
  \phi({\bf w}_{l+1}, {\bf p}_{l+1}) \le  \phi({\bf w}_{l}, {\bf p}_{l}) + \tau^N  \psi_l  \text{grad}^H(\phi({\bf w}_l, {\bf p}_l))  {\bf d}_l,
\label{dss}
\end{equation}
where $\tau \in (0,1)$ the searching coefficient; $N$ is the number of linear searches; $\psi_l$ is the initial step size. 

\subsubsection{Updating the next iteration point}
According to the search step size and descent direction, the next iteration point is updated as follows,
\begin{equation}
    {\bf w}_{l+1} = {\bf w}_{l} +  \tau^N  \psi_l {\bf d}_{{\bf w}_l}, \quad\quad
   {\bf p}_{l+1} = {\bf p}_{l} +  \tau^N  \psi_l {\bf d}_{{\bf p}_l}. \label{descentR}
\end{equation}
Since the update of (\ref{descentR}) may result in the next iteration point falling out of the product manifold $\mathcal{M}$, a retraction operation $\text{Ret}_{({\bf w},{\bf p})}({\bf w}_{l+1}, {\bf p}_{l+1})$ is necessary to map the resulting point back onto $\mathcal{M}$ to ensure feasibility, given as \cite{zhong2022mimo},
\begin{equation}
({\bf w}_{l+1},  {\bf p}_{l+1}) = \text{Ret}_{({\bf w},{\bf p})}({\bf w}_{l+1},  {\bf p}_{l+1}) = [{\bf w}_{l+1} \oslash {\bf w}_{l+1};{\bf p}_{l+1}].
\label{Rectrection}
\end{equation}
where $\oslash$ denotes elementwise division. 

In summary, the PCGD algorithm for solving (\ref{reforoverpro}) is presented in Algorithm \ref{alg:1}.

\begin{algorithm}
	\floatname{algorithm}{Algorithm}
	\renewcommand{\algorithmicrequire}{\textbf{Input:}}
	\renewcommand{\algorithmicensure}{\textbf{Output:}}
	\caption{: The PCGD algorithm to the problem (\ref{reforoverpro}).}
	\label{alg:1}
	\begin{algorithmic}[1]
		\REQUIRE 
                ${\bf w}_j, {\bf p}_j,\varepsilon_{i} $.\\
            \STATE Initialize $l = 0$, ${\bf w}_l={\bf w}_j$ and ${\bf p}_l={\bf p}_j$.   
            \STATE {\textbf {Reapeat}}
            \STATE \quad Calculate the descent direction ${\bf d}_l$ by (\ref{Calculationdescent});
            \STATE \quad Calculate the descending step size $\tau^N  \psi_l$ by (\ref{dss});
            \STATE \quad Updating the next iteration point ${\bf w}_{l+1},{\bf p}_{l+1}$ by (\ref{Rectrection});
            \STATE \quad $l \leftarrow l+1$;
		\STATE {\textbf {Until } $|\phi({\bf w}_{l+1}, {\bf p}_{l+1})-\phi({\bf w}_l, {\bf p}_l)|<\varepsilon_{i}$.}\\
	\end{algorithmic}%
\end{algorithm}

\subsection{Update the Penalty Parameter}
The penalty parameter $\rho$ plays a role in constraining and balancing solutions in optimization problems. Intuitively, when the obtained point is far from feasible, the penalty parameter may be too small so that it should be increased to $\rho=\rho/c_1$, where $c_1\in(0,1)$.

In summary, the PPM method for solving (\ref{objectivefunction}) is presented in Algorithm \ref{alg:2}. The parameter $\sigma$ represents the maximum violation of the inequality constraints. If $\sigma > \eta$, Algorithm \ref{alg:2} adjusts the penalty parameter to enforce the inequality constraints obtained; otherwise, it keeps the penalty parameter unchanged for the next iteration. Algorithm \ref{alg:2} stops once $\sigma \le \sigma_{\min}$. We set $c_2$ within the interval $(0, 1)$ to ensure convergence, which are employed to iteratively decrease $\eta$.

\begin{algorithm}
	\floatname{algorithm}{Algorithm}
	\renewcommand{\algorithmicrequire}{\textbf{Input:}}
	\renewcommand{\algorithmicensure}{\textbf{Output:}}
	\caption{: The PPM method to the problem (\ref{objectivefunction}).}
	\label{alg:2}
	\begin{algorithmic}[1]
		\REQUIRE 
                Initialize $ {\bf w}_0, {\bf p}_0,\rho_0,\sigma_0 ,\sigma_{\min},\eta_0,\varepsilon_{o}, j = 0$.\\
            \STATE {\textbf {Reapeat}}
            \STATE \quad Update ${\bf w}_{j+1},{\bf p}_{j+1}$ by Algorithm \ref{alg:1};
            \STATE \quad $\sigma_{j+1} = \max\left\{\begin{array}{l} \max\{0,g_m(p_{m+1},p_{m})\},\forall m, \\ \max\{0,f(p_{1})\},\max\{0,f(p_{M})\}  \end{array} \right\}$;
            \STATE  \quad\textbf{IF} $\sigma_{j+1} >\eta_{j} $
            \STATE \quad\quad $\rho_{j+1}=\rho_{j}/c_1$; 
            \STATE  \quad\textbf{ELSE}
            \STATE \quad\quad $\rho_{j+1}=\rho_{j}$;
            \STATE  \quad\textbf{END IF}
            \STATE \quad  $\eta_{j+1} = c_2 \cdot \sigma_{j+1}$, $j=j+1 $;
		\STATE {\textbf {Until } $\sigma_{j+1}  \le \sigma_{\min} $ and $||[{\bf w}_{j+1},{\bf p}_{j+1}]-[{\bf w}_{j},{\bf p}_{j}]||_2^2<\varepsilon_{o}$.}\\
	\end{algorithmic}%
\end{algorithm}

\subsection{Analysis of Complexity and Convergence}
The primary complexity of the PPM method in Algorithm \ref{alg:2} lies in updating ${\bf w}_{j+1}$ and ${\bf p}_{j+1}$ using Algorithm \ref{alg:1}. The complexity of Algorithm \ref{alg:1} is primarily attributed to the computation of the Euclidean gradient $\nabla \phi({\bf w}, {\bf p})$ by (\ref{updategradient}), which is about $\mathcal{O}(M^2)$. For the convergence, in the inner loop of Algorithm \ref{alg:1}, the Armijo backtracking line search algorithm is used to determine the step size, while the Polak-Ribiere parameter is employed to compute the conjugate direction, ensuring that the objective function does not increase at each iteration \cite{babaie2015hybridization}. Additionally, the objective function is smooth with a lower bound of 0. Based on this, we can ensure that every limit point generated by the algorithm is a critical point, as established in \textit{Proposition 4.7} of \cite{boumal2023intromanifolds}. Since Algorithm \ref{alg:1} converges to a critical point, we further show that every limit point of the iterates generated by Algorithm \ref{alg:2} is a KKT point of problem (\ref{objectivefunction}) when $\sigma_{\min} = 0$. Convergence is guaranteed by verifying the conditions outlined in \textit{Proposition 4.2} of \cite{liu2020simple}.

\begin{figure}[t]
  \begin{center}
  \includegraphics[width=2.5in]{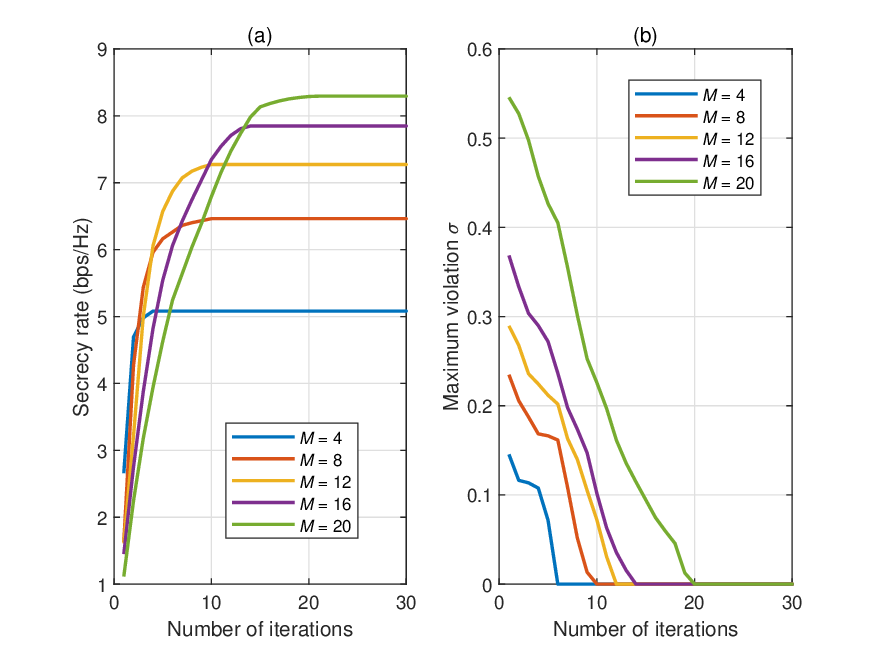}\\
  \caption{Convergence of the proposed PPM method.}\label{Fig1}
  \end{center}
  \vspace{-0.6cm}
\end{figure}

\begin{figure}[t]
  \begin{center}
  \includegraphics[width=2in]{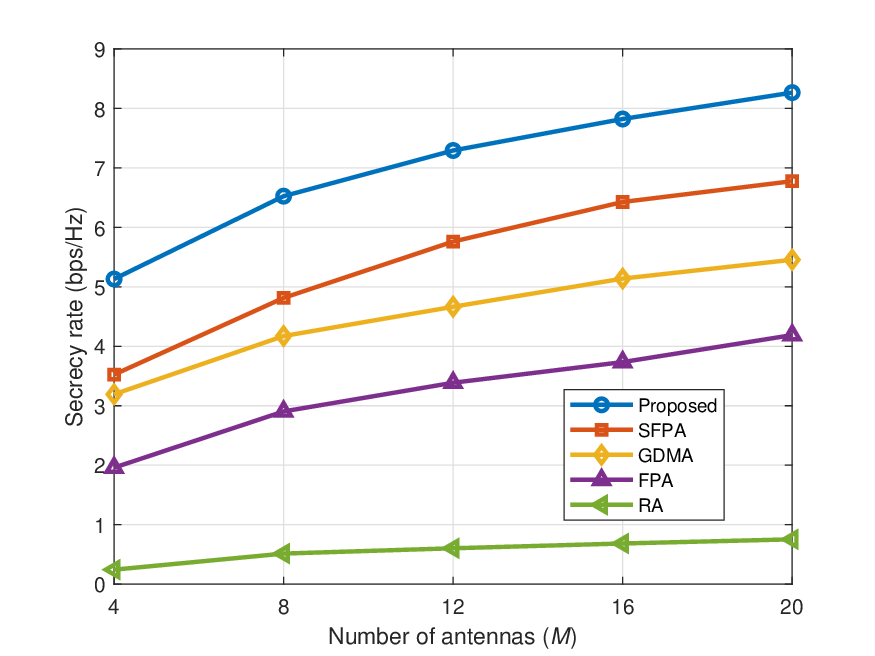}\\
  \caption{Secrecy rate with different numbers of antennas at Alice ($M$).}\label{Fig2}
  \end{center}
  \vspace{-0.6cm}
\end{figure}

\begin{figure}[t]
  \begin{center}
  \includegraphics[width=2in]{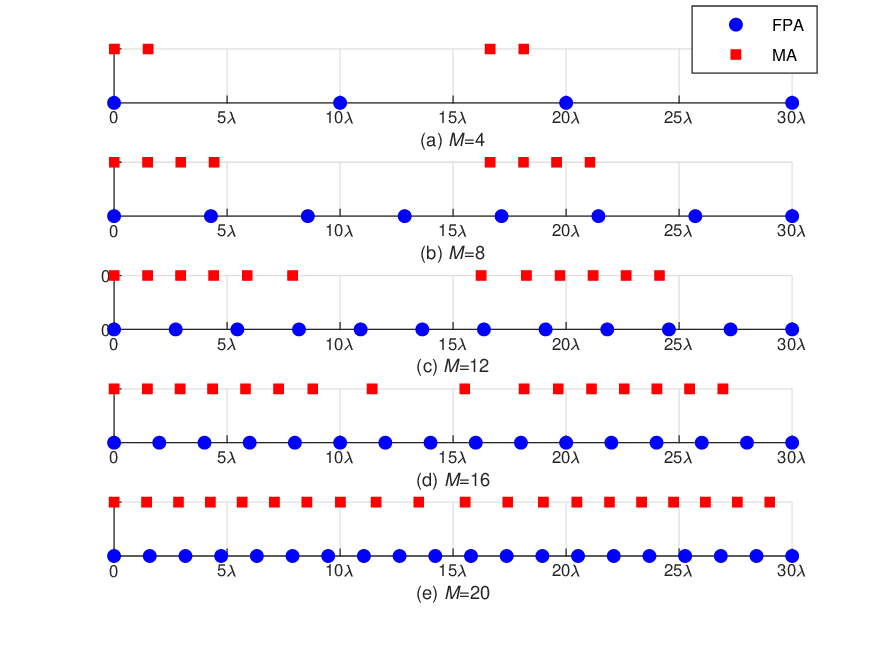}\\
  \caption{The antenna positions for the MA and the FPA.}\label{Fig3}
  \end{center}
    \vspace{-0.6cm}
\end{figure}

\section{Numerical Results}
In this section, we compare the proposed method with the following benchmark schemes: 1) \textbf{GDMA} \cite{hu2024secure,cheng2024enabling}: the gradient descent method, used to enhance PLS in the FDB MA system, is applied here with direct projection to achieve constant modulus; 2) \textbf{SFPA}: the FPA system selects antennas to form a sparse array \cite{bose2021efficient}; 3) \textbf{FPA}: the FPA system with a uniform linear array (ULA) \cite{li2019constant}; and 4) \textbf{RA}: the system with randomly selected antenna positions. In simulation, we use the settings $\lambda=0.01$m, ${\sigma_b^2}=1$, ${\sigma_e^2}=1$, $P_k=1$W, $L = 30\lambda$, $L_b=L_e=4$, $\beta_{i,l} \sim \mathcal{C N}({0}, \frac{1}{L_i})$ for $l=1,...,L_i$ and $i \in \{b,e\}$, and azimuth angles are randomly set within $[0,\pi]$.

Fig. \ref{Fig1} shows the secrecy rate as a function of the number of iterations for Algorithm \ref{alg:2} with different MA array sizes ($M$). As shown in Fig. \ref{Fig1}(a), a smaller $M$ leads to faster convergence due to reduced data dimensionality, while a larger $M$ achieves a higher secrecy rate at convergence, benefiting from an increased DoF. All cases converge within 20 iterations, demonstrating the method's fast convergence. Fig. \ref{Fig1}(b) shows that the maximum violation $\sigma$ decreases monotonically to 0, indicating that the antenna position constraints are satisfied by adjusting the parameter $\rho$ in Algorithm \ref{alg:2}.

Fig. \ref{Fig2} illustrates the secrecy rates for different architectures. The results show that the proposed method with the MA array achieves a higher secrecy rate compared to the SFPA, FPA, and RA architectures. Moreover, the proposed method with MA outperforms the GDMA method from \cite{hu2024secure,cheng2024enabling}, highlighting that the proposed approach is better suited for the AB system.

Fig. \ref{Fig3} illustrates the antenna positions of the proposed method and the FPA architecture for different numbers of antennas at Alice ($M$). The MA positions naturally form two clusters, which strategically optimize the channel information for Bob at specific angles. This clustering behavior arises because the MA system adjusts antenna positions to maximize the received signal strength at legitimate receivers, while minimizing interference from eavesdroppers. Unlike the FPA array, where antennas are distributed uniformly, the MA array's flexible positioning allows it to concentrate antennas in regions that enhance communication performance. This adaptive positioning results in a non-uniform distribution, improving the PLS.

\section{Conclusion}
 In this paper, we considered an analog beamforming system with MA. We aimed to maximize the secrecy rate by jointly designing the transmit beamforming and the positions of all antennas at the base station. A penalty product manifold method were proposed to solve the challenging non-convex problem effciently. Numerical results showed that the system with MA achieved a higher secrecy rate.

\bibliographystyle{IEEEbib}
\bibliography{strings,refs}

\end{document}